# Comparative Analysis on Inertia Estimation Algorithms (IEAs) in Providing Proper Frequency Response

**Karl M.H. LAI†, Yunhe HOU* and Kwunhang WONG****

**Abstract** – The inertia constant H[s] is a fundamental indicator of power system resilience, linking power imbalance between generation and load to frequency deviation. It is essential in frequency reserve dispatch under stability-constrained optimal power flow (OPF), demand response (DR) in ancillary service, system decoupling and frequency control in modern power system. While the inertia constant is traditionally defined as the intrinsic kinetic energy of synchronous generators on bar normalized to the power base, this neglects the releasable power under nonlinear dynamics and control inside HVDC and Inverter-based Resources (IBRs). Accurate real-time inertia estimation is therefore essential to perform proper frequency control and to indicate the risks of failure in frequency restoration. It, however, is challenging with noisy frequency measurement under event-driven parameter jumps and locational transient responses.

This paper presents a systematic comparative analysis on inertia estimation algorithms (IEAs) for frequency response applications. Classical methods such as filtering and fitting under measurement-based methods are benchmarked against data-based parameter estimation techniques such as recursive least squares (RLS) and model-based methods such as Kalman filtering (KF).

The main contributions are: (i) a holistic review of model- and data- based inertia estimation methods, (ii) exploration on the effect of IEA to wind-based inertia emulation strategies. The findings underscore the need for robust, adaptive, and data-driven estimation frameworks to ensure secure operation of future low-inertia grids.

## 1. Introduction – Frequency Disturbance and Inertia in Modern Power Systems

Power system frequency fs [Hz] is controlled by the power deviation between generation and load. Any power imbalance is a frequency disturbance possibly driving the system into abnormal operating ranges leading to equipment damage or even system failure. Typical frequency disturbances include load rejection, abrupt load steps, output power fluctuation in inverter-based resources (IBRs), generator tripping, HVDC shutdown and faults that reduce voltage and hence power transfer capability. The domination of synchronous generators in history provides significant inertia, which was an immediate buffer against sudden power imbalance, reducing rate of change of frequency (ROCOF) such that frequency activation and under frequency load shedding (UFLS) as system protection have time to response [1], [2]. With proper governor and automatic generator control (AGC) and plenty of spinning reserves (SR), frequency instability was not a big concern in the past.

The increasing infeed from IBR plants such as solar photovoltaics and wind turbines aided with battery energy storage systems (BESS), as illustrated in **Fig. 1**, has changed this condition. As IBRs are decoupled from grid frequency through grid side converters (GSCs), the IBRs do not necessarily contribute inertia. Even if they contribute to inertia or frequency response in some forms, such as virtual inertia [15], virtual synchronous generator (VSG) [13] droop control [16], or fast frequency response (FFR) in BESS, the actual response is not guaranteed with intermittent power input or limited state of charge (SOC) in the BESS.

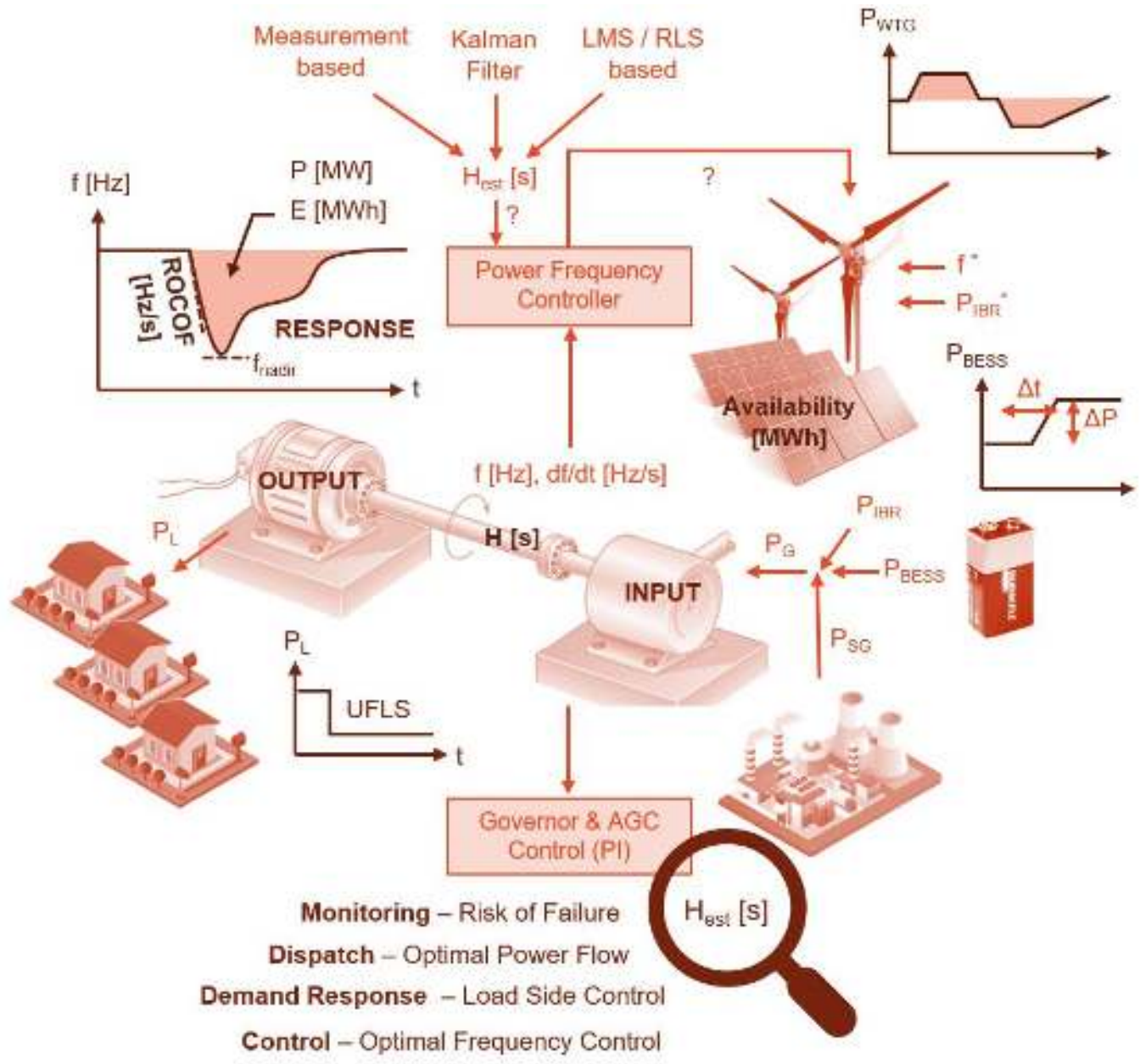

**Fig. 1.** Analogy for Frequency Control & Inertia Estimation

† Corresponding Author : Dept. of ECE, The University of Hong Kong & CLP Power Hong Kong (perspect@connect.hku.hk)

* Dept. of Electrical and Computer Engineering, University of Hong Kong (yhhou@eee.hku.hk)

** Dept. of Electrical and Computer Engineering, University of Hong Kong (u3556440@connect.hku.hk)

The consequence is a continuous decrease in system inertia in regions with significant use of IBRs and HVDCs import, causing concerns about elevated ROCOF with lower frequency nadirs, increasing risk of cascade tripping with UFLS failure due to measurement delays [2], which requires 2.5 cycles at least with phasor measurement unit (PMU) to measure an accurate ROCOF [17].

Recent operational experience indicates the need to provide real-time monitoring and control of system inertia. For example, ENTSOE in Europe [5] and POSOCO in India [7] initiated real time inertia monitoring under the declining spinning reserve on bar. Without direct monitoring, CAISO in California has documented a regional ROCOF of 0.422Hz/s, 6 times greater than the interconnection-wide averages under midday solar peak with dominating solar output, indicating the possibility of system collapse under large frequency disturbance with limited inertia for frequency arrest [4]. It necessitates the needs to have clear definition on a "*critical inertia*", as did in ERCOT in Texas with a threshold of 100 GWs to prevent UFLS [6] and procurement of inertia as an ancillary service in AEMO of Australia [3].

Inertia estimation in practice is often for dispatch and monitoring. While BESSs are often proposed as the major solution to frequency arrest, it is often found with insufficient SOC when needed. It can be mitigated with a proper output prediction and pre-charging. Another possible solution is a proper frequency response from IBRs, as a source with inertia extraction. To enable the capability, it intrinsically is a real-time decentralized control problem, and model-based approaches can provide proper frequency response with regards to system condition and energy availability [MWs] of IBR plants. The uncertainty parameters are the varying inertia constants and the energy availability, such as wind speed or state-of-charge. As a result, precise inertia estimation is essential for inertia extraction.

Inertia estimation, however, is still challenging even under such urgent need. As discussed, the parameters currently obtained for inertia monitoring ranges from inertia constant H[s], available energy [MWs] or even ROCOF [Hz/s]. Traditional theory implies that inertia constant is the available kinetic energy normalized by the based power and it can be calculated with the parameters for generators on bar as follows.

$$H_{eq} = -\frac{\Delta P}{2\frac{d\left(\frac{\Delta f}{f_0}\right)}{dt}\Big|_{t=0^+}} = \frac{\left(\overbrace{\sum_{i=1}^{CP} H_i\, S_{B,i}}^{H_{R,eq}\,S_{G,B}} + \overbrace{\sum_{j=1}^{VG} H_{V,j} S_{B,j}}^{H_{V,eq}\,S_{IBR,B}}\right)}{S_B} \quad (1)$$

where $\Delta P$, $\Delta f$, $H_i$, $H_{V,j}$ and $S_B$ are power deviation, frequency deviation, inertia constant for generator $i$, virtual inertia parameter used in IBR $j$ and based power [MVA] respectively.

The parameter $H_{V,j}$ (V = virtual inertia) is not a given parameter, since the settings used for VSG or droop control with front-end controllers may not directly correspond to the actual output governed by the outer power control loop. A measurement-based approach, therefore, is often required for inertia estimation. Further challenges include frequency estimation as pre-processing, real time power imbalance calculation, event start-time identification, ROCOF denoising and locational effect of measurements.

Academic and operational communities have proposed diverse methodologies for inertia estimation, ranging from generator parameter summations to event-driven analyses using disturbance data. Each approach, however, faces limitations in accuracy, scalability, and applicability under high renewable penetration.

The inertia constant H is often estimated based on the swing equation [18] –

$$2H\frac{df'}{dt} + Df' = \underbrace{P_G - P_L}_{\Delta P}, \qquad f' = f_s - f_n \quad (2a)$$

$$f'(t) = \left(\Delta f_0 - \frac{\Delta P}{D}\right)e^{-\frac{D}{2H}t} + \frac{\Delta P}{D} \quad (2b)$$

$$\frac{df'}{dt}(t) = -\frac{D}{2H}\left(\Delta f_0 - \frac{\Delta P}{D}\right)e^{-\frac{D}{2H}t} \quad (2c)$$

where $D$ is the damping constant and $f'$ is the shifted frequency in center of inertia (COI).

Inertia estimation algorithm (IEA) based on the swing equation ranging from measurement-based approach supported with fitting, filtering and robust statistics, Kalman Filtering (KF) and parameter estimation solution with least mean square (LMS) and recursive least square (RLS) approach are evaluated in this paper.

Observability of real-time inertia is essential for frequency reserve coordination under stability-constrained optimal power flow (SC-OPF), where dispatch decisions must respect different stability limits [9], [10]. Inertia estimates also inform demand response (DR) participation in ancillary services, ensuring that activation thresholds and energy budgets are aligned with instantaneous system resilience [11], [12]. UFLS schemes and system decoupling protections should be coordinated with frequency control activation and adapted to the inertia metrics to avoid unnecessary dis-connection to prevent system collapse [10], [12].

The remainder of this paper is organized into six sections, each addressing a critical dimension of inertia estimation and its roles in modern power systems. **Section 2** introduces measurement-based approaches to determine the inertia constant with ROCOF based on the swing equation. **Section 3** focuses on model-based approach including square root Kalman Filter (SR-KF), Information Filter (IF), Adaptive Unscented Kalman Filter (AUKF) and Multi-model Adaptive Estimation (MMAE), emphasizing their ability to handle noisy frequency measurements with parameter jumps based on a frequency model. **Section 4** discusses data-driven parameter estimation techniques, particularly least mean squares (LMS) and recursive least squares (RLS), which offer adaptive solutions under variable operating conditions. **Section 5** evaluates the performance of inertia estimation in frequency control, with special attention to wind-based inertia emulation strategies. Finally, **Section 6** provides a summary of findings and outlines future directions toward robust, adaptive, and measurement-driven inertia estimation frameworks.

## 2. Inertia Estimation – Measurement-based Approach

The inertia constant traditionally was evaluated simply by considering the weighted average of that of the synchronous generators in the system, as indicated in the first term of (1), neglecting the damping effect of different load types, Primary Frequency Response (PFR) of the generators and motor inertia [1]. Yet, with the introduction of power-electronic devices such as HVDC infeed and IBR penetration, the inertial performance under the nonlinear control energy availability must be evaluated in real time, possibly for monitoring, dispatch, activation and control [2]-[12].

The first research for inertia estimation with measurement approaches was given by Inoue in 1997, with a focus on polynomial approximation (fifth order) to filter out the noise generated under average frequency measurement and ROCOF calculation, based on a noisy and small signal with a differentiator [19]. It uses the linear term constant to represent the ROCOF value assuming the nonlinear terms in local approximation are relatively small. The study assumed disturbance with capacity of 20% to 40% of loads lasted for 14 – 18 secs. It also indicated the need for Detrended Fluctuation Analysis (DFA) to determine event start time.

To avoid the amplification of noise from differentiators for ROCOF calculation, Wall and Terzija (2014) suggested a sliding window approach in which both the moving average of power deviation $\bar{P}_i(t_i)$ and frequency $\bar{\bar{f}}_i(t_i)$ were calculated within the window and the inertia constant was then estimated with the difference in power deviation and frequency between two windows as in (3) [23]. It was declared as a real-time measurement to detect the event start time $t$ with a cumulative sum CSUM within window length $A$ and to reject false measurement with a threshold $tr$. However, selection of these two parameters depends on system experience, noise level and size of disturbance. A dynamic clipping on $H \in [H_{\min}(t), H_{\max}(t)]$ was employed to ensure convergence. The result indicated that only a large enough disturbance indicated by the CSUM can provide a true estimate of H. It is noted that by this approach, it can reach a higher numerical stability in case different sign of $\bar{\bar{f}}_1(t_1)$ and $\bar{\bar{f}}_2(t_2)$ is selected.

$$\hat{H}(t;t_1,t_2) = \frac{1}{2}\frac{(\bar{P}_1(t_1) - \bar{P}_2(t_2))}{\bar{\bar{f}}_1(t_1) - \bar{\bar{f}}_2(t_2)} \tag{3}$$

The simple inertia estimation from Inoue to detect a proper ROCOF and the power deviation at the change period for approximation purposes is regardless of load effect with changing frequency and voltage. Zografos, from 2016 – 2018, has introduced the famous R-, V- and RV-method (4) by including additional terms to compensate for the load effect under frequency and voltage deviation respectively. [20] – [22]. The study was performed in Nordic system and assuming a significant power deviation upon generator losses and a sole event leading to such frequency deviation in the period. It evaluated the ROCOF in the linear region and quadratic region with polynomial fitting and low pass filtering (LPF – Butterworth) and compared with the performance with the use of lowest ROCOF to avoid underestimating the inertia constant. It was found that the estimation error could reach a mean of 5.48% with a variance of 46.60% [22].

$$\hat{H} = \frac{\left(\underbrace{R(t)|_{t=t_s}\,\Delta f|_{t=t_s}}_{\text{R-Method}} - \underbrace{P_{L0}(U_s(t)-1)}_{\text{V-Method}} - \Delta P_d\right)}{2\frac{df}{dt}|_{t=t_s}} \tag{4}$$

In the comparative analysis, the system frequency change was driven by random load steps with variance of 0.008pu in every 2 seconds to indicate a power deviation and a step-change in inertia value, possibly due to generator dispatch or different operation mode of IBRs and HVDCs. The additional output from generator and IBRs droop control is absorbed into the load steps, assuming the droop constant is small enough to create negligible slope at the load step. Additional Gaussian noise with S.D. = 0.001 is added to the frequency measurement to test the noise rejection capability.

To eliminate the noise effect in frequency and ROCOF measurement, exponential filter (Exp Filter), polynomial fit (Poly Fit), Savitzky-Golay (SG) smoothing and Butterworth LPF were compared. The ROCOF value was also filtered with median filter, as a robust statistical filter, and average value of all filters was tested to suppress the differentiator noise amplification and compared with the Inoue method discussed. A simple clipping with $\hat{H} \in (2.2, 3.5)$ was also used to avoid numerical instability under no excitation period, which has insignificant power deviation $dP$ and frequency shift $f'$. The formulations for the filters applied on frequency to reduce noise are as follows.

**Table 1.** Formulations for Filtering and Fitting

| Filter | Formulation |
|---|---|
| Exponential Filter ($\alpha$) | $y[n+1] = \alpha y[n] + (1-\alpha)x[n]$, $\alpha \in [0,1]$ (forgetting factor) |
| Butterworth LPF $(\omega_c, n)$ | $\lvert H(j\omega)\rvert = \dfrac{1}{\sqrt{1+(\omega/\omega_c)^{2n}}}$ |
| Polynomial Fit ($n$) | $\hat{\mathbf{a}} = \min_{\mathbf{a}} \sum_{i=1}^{M} (y_i - \mathbf{a}^{\mathrm{T}}\mathbf{x}_i)^2$, $\hat{y} = \hat{\mathbf{a}}^{\mathrm{T}}\mathbf{x}_i$ where $\mathbf{x}_i = \left[1\ x_i\ x_i^2 \dots x_i^N\right]^{\mathrm{T}}$ (Global fit to all data set in the window) |
| SG Filter [24] | $\min\lVert\mathbf{y} - \mathbf{Ah}\rVert^2$ $p_0 = a_0 = \mathbf{h}^{\mathrm{T}}\mathbf{y} = (A(A^TA)^{-1}e_1)^{\mathrm{T}}y$ where $y = [y_{-m}\ y_{-m+1} \cdots y_0 \cdots y_{m-1}\ y_m]^{\mathrm{T}}$, $(A)_{ij} = (i-1-m)^{j-1}$ is the Vandermonde matrix, and $e_1 = [1\ \mathbf{0}^{\mathrm{T}}]^{\mathrm{T}}$ is the basis vector to return the local constant. (Local fit to the current point in the window) |

**Table 2.** Performance Index for Inertia Estimation - Measurement based Approach

| Method | $\mathbb{E}\left\lVert\dot{f} - \hat{\dot{f}}\right\rVert^2$ | $\mathbb{E}\lVert H - \hat{H}\rVert^2$ | $\bar{\hat{H}}$# |
|---|---|---|---|
| Exp Filter | $7.76 \times 10^{-6}$ | 0.0308 | 2.581 |
| PolyFit | $2.66 \times 10^{-6}$ | **0.0206** | **2.555** |
| SG | $2.15 \times 10^{-6}$ | 0.0228 | 2.550 |
| LPF | $1.89 \times 10^{-6}$ | 0.0219 | 2.548 |
| Median | $2.19 \times 10^{-6}$ | 0.0218 | 2.573 |
| Average | $\mathbf{1.39 \times 10^{-6}}$ | 0.0213 | 2.572 |

# The true average value of H over the 200s is 2.560.

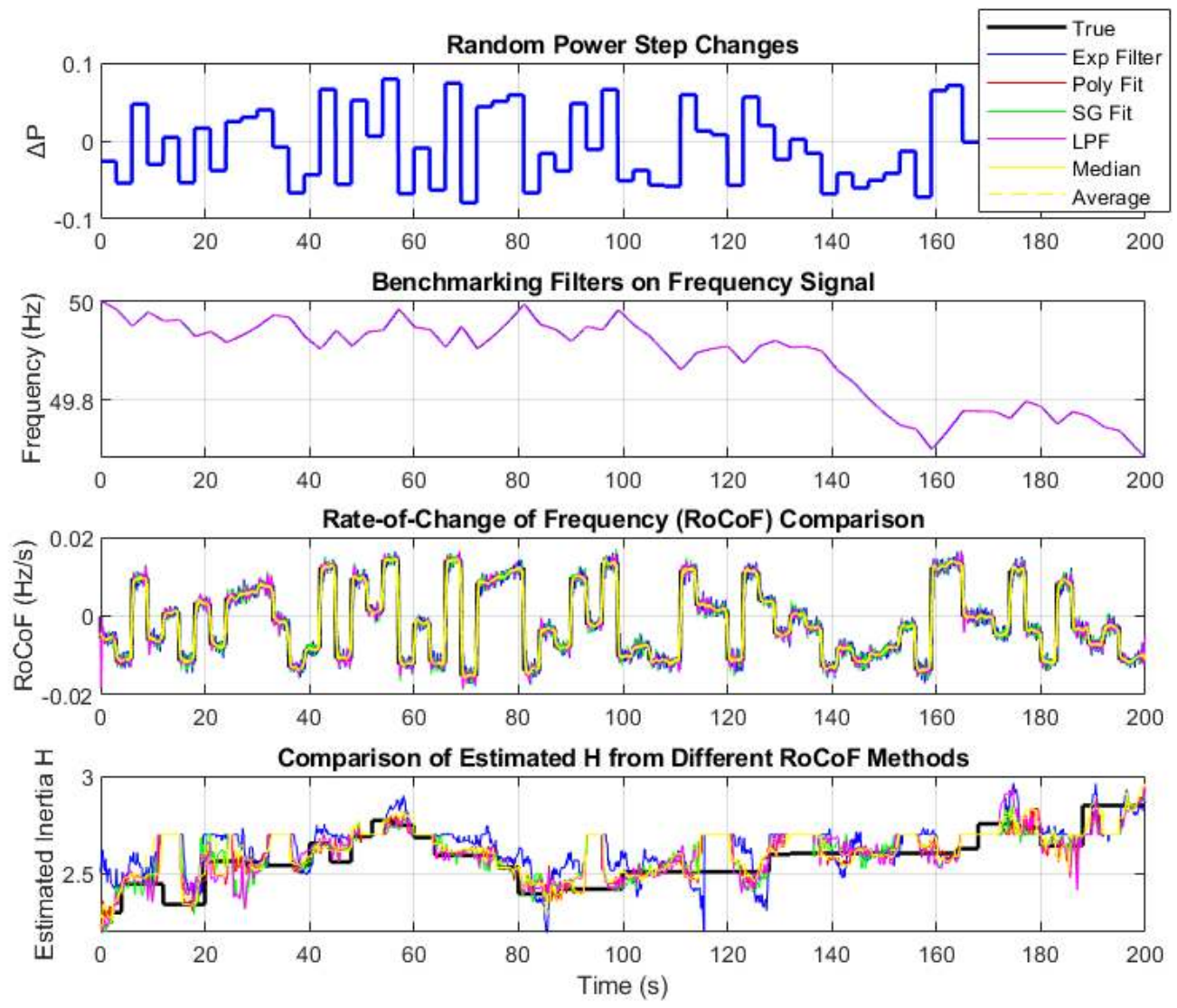

**Fig. 2.** Estimated Inertia with Measurement-based Method

From the simulation, the performance of ROCOF calculation and inertia estimation upon general load steps in real time, which is not the same as in literature. From **Figure 2**, the noise of frequency signal was filtered with the respective filter or fitting methods to generate the ROCOF values, and the ROCOF values were then used to estimate the inertia value. It was noted from the last plot on estimated inertia that the measurement-based method does not track well with a varying H, even the mean square error (MSE) as listed in **Table 2** for inertia estimation is in average of as the differentiator noise in ROCOF is amplified with the 1/ROCOF operation in the calculation. The situation becomes worse when the ROCOF value is around zero, when there is a sign changing load step. To avoid such conditions, one can consider calculating the inertia value with absolute value of power deviation and ROCOF. The performance can only be improved by keeping the previous value or returning to a default value, possibly the largest possible value to indicate a susceptible value or the typical value. From the performance index shown in **Table 2**, the estimated value of ROCOF is the best with averaging, with the estimated ROCOF of all methods are summed up and taking an average; and the estimated inertia has the smallest MSE with polynomial fitting in a sliding window. It is possibly due to the smoothening and delaying effects of the filters (LPF or Exp Filter) distorting the ROCOF value and the localized SG filter does not effectively filter out the noise. The time average of estimated inertia constant of all filters (2.548 ~ 2.581), however, is close to the true value (2.560), which indicates the possible application of inertia monitoring, where large sampling period (~ 200s, in order of minutes) is acceptable. It is noted that the inertia change is exaggerated in the simulation, as the mode change in IBR operation or generator tripping should not be as frequent. Also, from simulation result in practice, using minimum ROCOF over time as did in Zografos [20]-[22] for the estimation tends to overestimate the inertia constant, and using two-window averaging approach in Wall and Terzija [23] often create corresponding fluctuations over time without effective start-time identification or false measurement rejection.

## 3. Inertia Estimation – Model-based Approach

Model-based approach, as compared to measurement-based approach, has the advantage of avoiding the ROCOF value to compute the inertia constant, such that the estimated inertia constant is more stable under small ROCOF value due to negligible load steps. Kalman Filter is an optimal state-observer which minimizes the error covariance with an optimal Kalman gain to compensate the estimation error upon measurement update. It assumes a dynamical system interfered by Gaussian process noise with algebraic measurement corrupted by Gaussian measurement noise. If the system dynamics is not dominant, the state follows a near random walk with the noise. Hence, the initial order to the diagonal element of the covariance matrix and the noise covariance should be properly sized to indicate the possible steps to be observed in the state for proper tracking.

Kalman Filter, other than serving as an optimal state observer, It can also be used as a parameter estimation tool, or to perform dual estimation, in different areas such as battery parameter [27]-[30], heat transfer parameter [31], and aviation parameter [32] with different techniques such as Extended Kalman Filter (EKF) [26], [31], Unscented Kalman Filter (UKF) [27]-[29] and a hybrid of them [32]. Fan (2013) treated the inertia estimation a dual estimation problem in which considered parameters as additional states such that the linear dynamics in generator (rotor angle $\delta$ and rotor speed $\dot{\delta}$) with inertia constant H, damping constant D, input mechanical power $P_m$ and subtransient reactance $x_d{}'$ became a nonlinear system with the unknown states [26]. The system evolution matrix A included the generator dynamics and a random walk of the parameters. Iterative EKF (IEKF) which re-linearized the system at each step was utilized to improve convergence and reduce linearization errors. It compares the cases with classical generator model with or without damping, a sub-transient model with damper winding and field dynamics and one with AVR enabled. It was found out that EKF was insufficient to track the unmodelled dynamics.

For comparative analysis, a standard Square-Root Kalman Filter (SR-KF), which is a numerically stable variant of the standard KF with Cholesky Factorization of covariance to be propagated, is used. The SR-KF, as described in **Algorithm 1**, improves numerical stability and reduces round-off error for the ill-conditioned parameter estimation problem [28], [33]. To avoid numerical instability with a non-positive definite $\mathbf{S}^{+}_{\hat{\theta},\mathbf{k}}$, the last step measurement update on covariance in SRKF is eliminated to avoid failure in performing Cholesky update. It coincides with the performance of recursive least square (RLS) algorithm to be discussed in **Section 4**, with no error covariance shrinkage over time. For inertia estimation, SRKF can avoid numerical instability with small signals for both frequency and power deviation.

Another elegant yet computationally heavier form is Information Filter (IF), which avoids computing Kalman Gain with inverse and performing additional regularization and biasing for the bad conditioned error covariance over cycles, is as illustrated in **Algorithm 2** for comparison. The formulation for the estimation problem is as follows.

Consider a parameter estimation problem with the swing equation discretized with Forward Euler approximation to avoid instability under conversion of log function, which return no value or complex value due to asymmetry with exact exponential discretization –

Discretized Dynamics -

$$f'[k+1] = \underbrace{\left(1 - \frac{D\Delta t}{2H}\right)}_{a_k} f'[k] + \underbrace{\frac{\Delta t}{2H}}_{b_k} \Delta P[k] \quad (5a)$$

State Evolution -

$$\boldsymbol{\theta}[k] \triangleq \begin{pmatrix} a_k \\ b_k \end{pmatrix} \rightarrow \boldsymbol{\theta}[k+1] = \mathbf{I}_2 \boldsymbol{\theta}[k] + \mathbf{v}[k] \quad (5b)$$

Measurement -

$$\underbrace{f'[k+1]}_{z[k]} = \underbrace{(f'[k] \quad \Delta P[k])}_{\boldsymbol{\Phi}[k]} \underbrace{\begin{pmatrix} a_k \\ b_k \end{pmatrix}}_{\boldsymbol{\theta}[k]} + w[k] \quad (5c)$$

$$\text{with } \mathbb{E}\{\mathbf{v}\} = \mathbf{0}, \mathbb{E}\{w\} = 0, \mathbb{E}\{\mathbf{v}\mathbf{v}'\} = \mathbf{Q}, \mathbb{E}\{ww'\} = \mathbf{R} \quad (5d)$$

where $\mathbf{v}[k]$ and $w[k]$ are process noise and measurement noise tackling jumps in parameters and errors in frequency measurement respectively.

It is noted that the sampling period in (5a) should be large enough, as a regularization by discretization [34], to avoid numerical instability to estimate two parameters in different orders. Also, state evolution equation (5b) is intrinsically a random walk equation, which is not intentionally to tackle a tracking problem with step change.

**Algorithm 1** – SRKF

**Initialization**

1. Set

$$\hat{\boldsymbol{\theta}}_0^+ = \mathbb{E}[\boldsymbol{\theta}_0]$$
$$\boldsymbol{\Sigma}^+_{\hat{\theta},0} = \mathbb{E}\left[(\boldsymbol{\theta}_0 - \hat{\boldsymbol{\theta}}_0^+)(\boldsymbol{\theta}_0 - \hat{\boldsymbol{\theta}}_0^+)^{\mathrm{T}}\right]$$
$$\mathbf{S}^+_{\hat{\theta},0} = \text{cholesky}(\boldsymbol{\Sigma}^+_{\hat{\theta},0}, \text{'lower'})$$
$$\mathbf{S}^+_{\tilde{v},0} = \text{cholesky}(\mathbf{Q}, \text{'lower'})$$
$$\mathbf{S}^+_{\tilde{w},0} = \text{cholesky}(\mathbf{R}, \text{'lower'})$$

**Time Update (for each time step k) -**

1. $\hat{\boldsymbol{\theta}}_k^- = \hat{\boldsymbol{\theta}}_{k-1}^+$
2. $\mathbf{S}^-_{\hat{\theta},k} = \text{cholupdate}\left(\left(\mathbf{S}^+_{\hat{\theta},k-1}\right)^T, \mathbf{S}^{\mathrm{T}}_{\tilde{w}}\right)^T.$

**Measurement Update (for each time step k) -**

3. $\hat{\mathbf{z}}_k = \boldsymbol{\Phi}_k \hat{\boldsymbol{\theta}}_k^-$
4. Kalman Gain Update -

$$\mathbf{S}_{\tilde{z},k} = \text{cholupdate}\left(\left(\boldsymbol{\Phi}_k \mathbf{S}^-_{\hat{\theta},k}\right)^{\mathrm{T}}, \mathbf{S}^{\mathrm{T}}_{\tilde{v}}\right)^{\mathrm{T}}$$
$$\mathbf{M}\mathbf{S}^{\mathrm{T}}_{\tilde{z},k} = \mathbf{S}^-_{\hat{\theta},k}\left(\mathbf{S}^-_{\hat{\theta},k}\right)^{\mathrm{T}} \boldsymbol{\Phi}_k^{\mathrm{T}}$$
$$\mathbf{L}_k \mathbf{S}_{\tilde{z},k} = \mathbf{M}$$

5. $\hat{\boldsymbol{\theta}}_k^+ = \hat{\boldsymbol{\theta}}_k^- + \mathbf{L}_k(\mathbf{z}_k - \hat{\mathbf{z}}_k)$

**Algorithm 2** – Information Filter

**Initialization**

1. Set initial state and covariance $\hat{\boldsymbol{\theta}}_0 = \mathbb{E}[\boldsymbol{\theta}_0]$ and $\boldsymbol{\Sigma}_{\hat{\theta},0} = \mathbb{E}\left[(\boldsymbol{\theta}_0 - \hat{\boldsymbol{\theta}}_0^+)(\boldsymbol{\theta}_0 - \hat{\boldsymbol{\theta}}_0^+)^{\mathrm{T}}\right]$
2. Compute Information matrix $\mathbf{Y}_0 = \boldsymbol{\Sigma}^{-1}_{\hat{\theta},0}$ and Information vector $\mathbf{y}_0 = \mathbf{Y}_0 \hat{\boldsymbol{\theta}}_0$

**For each time step k,**

**Prediction -**

1. $\mathbf{Y}_{k+1} = (\mathbf{Y}_k + \mathbf{Q})^{-1}$
2. $\mathbf{y}_{k+1} = \mathbf{Y}_{k+1} \hat{\boldsymbol{\theta}}_k$

**Measurement Update -**

3. $\mathbf{Y}_{k+1} \leftarrow \mathbf{Y}_{k+1} + \mathbf{H}^{\mathrm{T}} \mathbf{R}^{-1} \mathbf{H}$
4. $\mathbf{y}_{k+1} \leftarrow \mathbf{y}_{k+1} + \mathbf{H}^{\mathrm{T}} \mathbf{R}^{-1} \mathbf{z}_k$
5. $\hat{\boldsymbol{\theta}}_{k+1} = \mathbf{Y}^{-1}_{k+1} \mathbf{y}_{k+1}$

**Table 3** - Performance Index for Inertia Estimation - SRKF and Information Filter

| H(0) | Method | MSE Mean | MSE S.D. | Max Error Mean | Max Error S.D. | $\bar{H}$ Mean | Bias H=2.37 |
|---|---|---|---|---|---|---|---|
| **2.37** | SRKF | 0.015 | 0.003 | 0.700 | - | 2.356 | −0.014 |
| | IF | 0.018 | 0.007 | 0.700 | - | 2.366 | −0.004 |

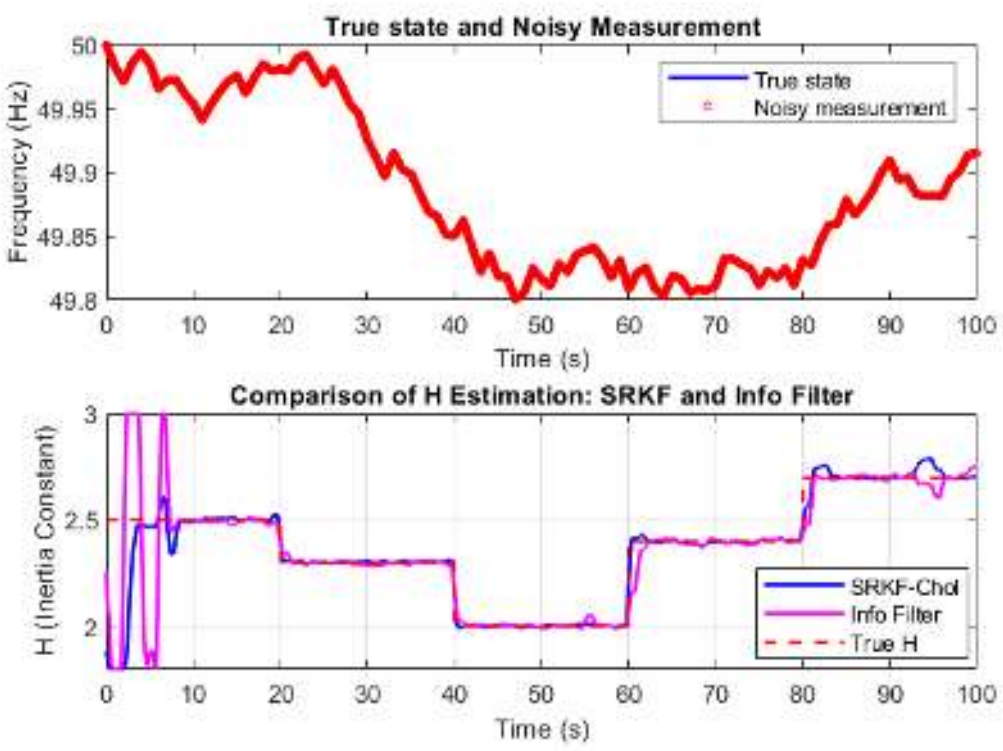

**Fig. 3.** Estimated Inertia with SRKF vs Info Filter

As illustrated in **Figure 3** and **Table 3**, it shows the strengths and weaknesses in SRKF and IF. SRKF achieves a lower MSE (0.015 vs 0.018) with a poorer result at the first few seconds of the filter initialization with the inverse of measurement noise covariance $R^{-1}$ to amplify the error until the filter converges. IF has a slightly higher MSE but lower bias. The larger bias in SRKF is due to the regularization steps or larger measurement noise to avoid failure in Cholesky factorization with a poor condition covariance matrix. The same result is as highlighted with the average estimated inertia over time, with IF mean estimate to be 2.366, closer to the true value. Also, IF contains a higher standard deviation in MSE, suggesting the sensitivity of the filter to noise as the filter is directly coupled to the measurement instead of the error itself. The maximum errors of both methods are the same due to clipping, which is active at filter initialization. It is noted that KF based approach generally performs better in inertia estimation as compared to the best (polynomial fitting) in measurement-based approach (MSE = 0.0206 in **Table 2**) as KF based approach does not amplify the noise with an improper use of small and noisy ROCOF value.

Another approach in KF based is by Unscented KF (UKF), a statistical-based observer introducing samples, known as sigma-points, to evolve over models to determine mean error and its covariance. UKF is often compared with EKF when the performance of EKF is questionable upon highly nonlinear model with the first order approximation. However, in parameter estimation with linear system, UKF often perform worse when process or measurement noise Q and R are mis-specified or intentionally to be tuned large to capture the step performance and it often results in unnecessary large covariance when the system is linear or mildly linear. For the inertia estimation computation, Adaptive UKF, as suggested in **Algorithm 3**, is used to improve the performance by amplifying the error covariance ($\Sigma_{\tilde{z},k} \leftarrow \alpha\Sigma_{\tilde{z},k}, \alpha > 1$) and the measurement noise covariance ($R \leftarrow R_l$ if $v_k > \varepsilon$) to capture the parameter jump due to generator trips or load shifts. To avoid the possible covariance shrinkage leading to non-positive definite covariance in calculation, Joseph Form covariance update in **Step 4** is used. It is noted that UKF intrinsically is still a KF with the parameter dynamics as a random walk which requires time to capture the jump. Also, the frequency estimation error is small even the inertia constant has a certain bias which indicates the colinear measurement and ill-conditioning of parameter estimation problem. The AUKF is also sensitive to initialization of error covariance to show the actual order of magnitude in each estimated parameter or state. Various adaptivity in [26]-[33] has been used to improve the performance.

Multi Model Adaptive Estimation (MMAE) is an alternative estimation technique based on EKF or UKF to determine the probability of the parameters in reaching the estimate by fixing the parameters in constant and retrieve the parameter estimate with the probability as in **Algorithm 4** [36]-[37]. It uses soft estimate in which returns the average instead of the one with maximum probability as maximum likelihood estimation (MLE) to avoid fluctuations. MMAE often avoids high dependence in initial guess and provides a result with interpretability; It, however, is much expensive in computation as compared to a pure UKF. It is noted that the performance should also be enhanced with the adaptive technique in **Algorithm 3**.

**Algorithm 3** – Adaptive Dual-UKF

**Initialization**

1. Define state vector $x = [f' \quad 1/H \quad D]^{\mathrm{T}}$.
2. Initialize estimate $\hat{x}_0 = [0 \quad 1/H_0 \quad D_0]^{\mathrm{T}}$ and covariance matrix $\Sigma_{x0} = \mathrm{diag}(\sigma_{f0}, \ \sigma_{1/H0}, \ \sigma_{D0})$
3. Choose the scaling factor $\alpha, \beta, \kappa$ and compute the weights $W_i^m, W_i^{(c)}$.

**For each time step k:**

1. Sigma Point Generation –
$$\chi_i(k-1) = \hat{x}^-(k-1) \pm \sqrt{(n+\lambda)P(k-1)}, \quad i = 0, \dots, 2n$$
2. Time Update –
$$\chi_i^-(k) = f(\chi_i(k-1), u_{k-1})$$
$$\hat{x}^-(k) = \textstyle\sum_i W_i^{(m)} \chi_i^-(k)$$
$$P^-(k) = \textstyle\sum_i W_i^{(m)} [\chi_i^-(k) - \hat{x}^-(k)][\chi_i^-(k) - \hat{x}^-(k)]^{\mathrm{T}} + Q$$
3. Measurement Update –
IF $[z(k) - \hat{z}(k)] > \varepsilon, \ R \leftarrow R_{\mathrm{Large}};$
ELSE $R \leftarrow R_{\mathrm{Small}};$
$$\zeta_i(k) = h(\chi_i^-(k))$$
$$P_{zz}(k) = \boldsymbol{\alpha} \textstyle\sum_i W_i^{(m)} [\zeta_i(k) - \hat{z}(k)][\zeta_i(k) - \hat{z}(k)]^{\mathrm{T}} + R$$
$$P_{xz}(k) = \textstyle\sum_i W_i^{(m)} [\zeta_i(k) - \hat{x}^-(k)][\zeta_i(k) - \hat{x}^-(k)]^{\mathrm{T}}$$

4. Kalman Gain and Update –

$$K(k) = P_{xz}(k)P_{zz}^{-1}(k)$$
$$\hat{x}(k) = \hat{x}^{-}(k) + K(k)[z(k) - \hat{z}(k)]$$
$$\hat{x}(k) \leftarrow \max(\min(\hat{x}(k), x_{min}), x_{max})$$
$$P(k) = (I - K(k)H)P^{-}(k)(I - K(k)H)^{\mathrm{T}} + K(k)RK^{\mathrm{T}}(k)$$

**Note –** The algorithm uses $x$ and $P$ to denote combined state and covariance to avoid mixing up symbols on covariance and summation.

**Algorithm 4** – MMAE

**Initialization**

1. Define candidate model $\{M_i\}_{i=1}^{I}$ including $H \in [H_{\min}, H_{\max}]$ and $D \in [D_{\min}, D_{\max}]$.
2. Initialize prior probability $p_i(0) = 1/K$ and UKF state estimate $\hat{x}_i(0)$ and $P_i(0)$.

**For each time step k:**

1. UKF Prediction (per model $M_i$) -
   a) Generate Sigma Points from $\hat{x}_i(k-1), P_i(k-1)$.
   b) Propagate sigma points through nonlinear dynamics.
   c) Compute predicted mean and covariance $\hat{x}_i^{-}(k), P_i^{-}(k)$.
2. UKF Update (per model $M_i$) -
   a) Transform Sigma Points through measurement model.
   b) Compute predicted measurement mean and covariance.
   c) Compute Kalman Gain $K_i(k)$ and update $\hat{x}_i(k), P_i(k)$.
3. Likelihood and Probability Calculation –

$$f_i(k) = \frac{\exp\left(-\frac{1}{2}v_i^T(k)P_{zz,i}(k)v_i(k)\right)}{\sqrt{(2\pi)^m |P_{zz,i}(k)|}}$$
$$p_i(k) = \frac{f_i(k)p_i(k-1)}{\sum_{i=1}^{I} f_i(k)p_i(k-1)}$$

4. Estimated Result –

$$\hat{x}(k) = \textstyle\sum_{i=1}^{I} p_i(k)\,\hat{x}_i(k);\ \ \hat{\theta}(k) = \sum_{i=1}^{I} p_i(k)\,\hat{\theta}_i(k);$$

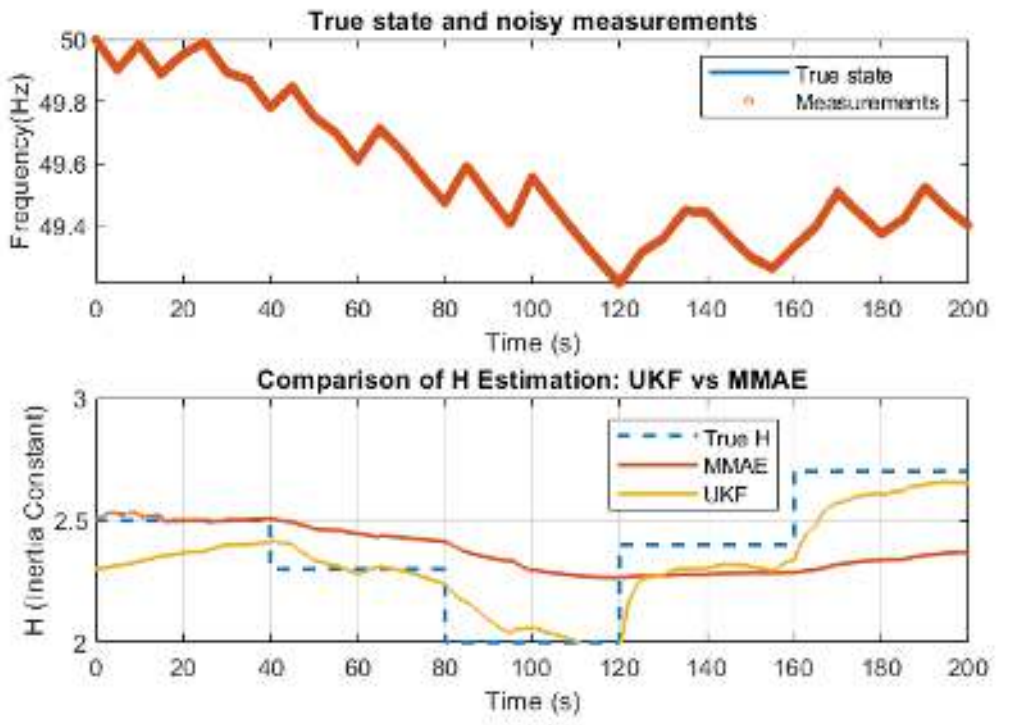

**Fig. 4.** Estimated Inertia with AD-UKF vs MMAE

**Table 4** - Performance Index for Inertia Estimation - AD-UKF and MMAE

| | | MSE | | Max Error | | $\bar{H}$ | Bias |
|---|---|---|---|---|---|---|---|
| **H(0)** | **Method** | **Mean** | **S.D.** | **Mean** | **S.D.** | **Mean** | **H=2.37** |
| **2.7** | ADUKF | 0.081 | 0.022 | 0.533 | 0.055 | 2.450 | +0.070 |
| | MMAE | 0.062 | 0.010 | 0.439 | 0.032 | 2.393 | +0.013 |
| **2.8** | ADUKF | 0.085 | 0.015 | 0.548 | 0.032 | 2.476 | +0.096 |
| | MMAE | 0.059 | 0.010 | 0.439 | 0.032 | 2.395 | +0.015 |
| **2.3** | ADUKF | 0.039 | 0.020 | 0.483 | 0.078 | 2.267 | −0.113 |
| | MMAE | 0.059 | 0.010 | 0.441 | 0.032 | 2.397 | +0.017 |

As illustrated in **Figure 4** and **Table 4** for the performance of AD-UKF and MMAE in inertia estimation, the MSE (~ 0.08 in UKF vs 0.02 in PolyFit) and the maximum error is worse than that of measurement-based technique even with adaptivity applied in UKF estimation. AD-UKF and further MMAE estimator often perform worse than the standard KF techniques when the system as a linear or mildly linear, and the process noise and error covariance are tuned large to capture jumps in parameter. However, in terms of robustness, the estimate seldom touches the clip bound, especially when MMAE intrinsically avoid such conditions. In the table, UKF's maximum error is more sensitive to initial guess; With a proper initial guess (H(0) = 2.3), the MSE can drop from 0.085 to 0.039.

AD-UKF is more precise when it is initialized near the true parameter, but it is prone to bias and larger MSE, while MMAE is more robust to poor initialization with lower bias and smaller maximum error, with its MSE slightly higher when the initial guess is already closed due to smoothening effect and with limited performance for fixed parameter state observer logic as compared to the actual dynamics. In practice, it is suggested to use UKF when prior knowledge of parameter is highly reliable as it is highly dependent on initial condition with proper parameter tuning on the spread of sigma point, or to use MMAE when robustness is more preferred in inertia estimation for control in power system.

In sum, the performance of SR-KF and IF in parameter estimation is much reliable and robust on initialization as compared to AD-UKF and measurement techniques. The SRKF based on linear parameter estimation does not require clipping and smoothing techniques and create large fluctuation as does in the measurement technique. The SRKF does not require tuning on initial conditions, filter parameters and adaptive parameters as in UKF to create a robust solution.

## 4. Inertia Estimation – Parameter Estimation Approach (LMS vs RLS)

Least Mean Square (LMS) algorithm and Recursive Least Square (RLS) algorithm are often the first choice in parameter estimation [38]-[40]. Wang (2023) employs weighed least square (WLS) approach, which is a variation to include a weight matrix, possibly with its diagonal elements corresponding to the error covariance of the estimated parameter, to minimize the weighed 2-norm of the residual. It stacks over the n-samples to create a tall matrix such that the three parameters to be estimated, namely [H, D,

R(Pm)], coincides with the RV formulation before. The underfit solution can help improve accuracy by avoiding too many colinear measurements especially when damping effect is insignificant such that the column rank is deficient. The performance of LMS is notable when the parameter estimated varies slowly, as LMS itself relies on (stochastic) gradient descent with step $\mu$ such that the algorithm goes towards the solution at the step size according to the local gradient. A small $\mu$ leads to a sluggish performance, and a large $\mu$ leads to oscillatory output. It has a slow adaptation to sudden change in estimate without an adaptive step size or a self-tuned algorithm. It is also sensitive to the scaling of the regressor [41]. Normalized LMS (NLMS) in **Algorithm 5**, includes a scaling factor [1 dt/2H] to reduce sensitivity of the algorithm to scaling for the two parameters, it also includes a denominator which contains the regularization term to avoid division of small numbers when the system has no excitation. The sluggish behaviour of LMS algorithm with a small step can possibly smoothen out the error and provide a stable and smooth inertia estimation for power system decisioning.

RLS, however, favoured for fast tracking of parameters when the parameters changed abruptly. It, however, requires a proper tuning for the forgetting factor $\lambda$, as a larger $\lambda$ leads to a slower tracking by penalizing less on the current error, while a smaller $\lambda$ leads to a faster tracking possibly with more peaky fluctuations. The algorithm is possibly destabilized by numerical errors, and it relies on a persistent excitation to track a proper $\theta$ [41]. Wang (2025) assumed an input-output model and used a dynamic optimal forgetting factor (DOFF)-RLS which minimized the sum of weighted residual to return an estimate [38]. The algorithm was used to estimate the inertia constant at different locations, and it was found to be more reliable with the synchronous generators than the wind farms. The RLS in **Algorithm 6** used is re-formulated to estimate the parameter 1/H instead to avoid division of small numbers to maintain numerical stability and the error covariance is initialized accordingly to reflect the estimate order of magnitude to prevent one parameter from dominating the update and after all the inertia estimate was clipped to suppress outlier and prevent divergence. It is noted that for control purposes, the RLS should be updated consistently whenever it has persistent excitation, and the estimate should serve for future activation purposes, as illustrated in **Section 5**.

As illustrated from the statistics in **Table 5** and **Figure 5**, LMS has a relatively high MSE (0.042 – 0.070) with large variation (±0.030 – 0.035), while RLS consistently achieves a much smaller MSE (0.009 – 0.012) with smaller variation (±0.003 – 0.004). The maximum error of RLS is much smaller (~ 0.39) upon parameter jumps. RLS also results in a closer average for inertia constant with near zero bias (-0.03 to +0.00) which is unbiased, while LMS shows a negative bias (-0.04 to -0.1) which underestimates the inertia constant which possibly leads to an overestimate of the ROCOF such that a model-based controller can provide an overshooting output. Also, both RLS and LMS algorithms create a delay in the parameter estimation as it requires enough samples to excite a proper frequency change and learn the parameters; RLS and LMS also intrinsically have a delay to either forget the previous learning or to walk in small steps towards to minimum descent point.

In sum, RLS outperforms other algorithms such as measurement-based approach, which is poorly performed due to the ROCOF calculation, and complex algorithms such as ADUKF and MMAE which are more favored to highly nonlinear environments. SR-KF and IF are robust, though with a slightly larger error than RLS, which requires a proper tuning of forgetting factors and initialization to avoid instability or peaky measurement.

It is noted that all algorithms, even SR-KF or RLS, do not perform well without persistent excitation, i.e. no power deviation and hence no frequency deviation. In case the inertia estimate is designated for frequency control and activation purposes, it is suggested to have a proper time coordination for the estimation process with the control. Controllers, which possibly remove all the ripples created by small power ripples, should be rendered inoperative until the frequency change is large enough.

**Algorithm 5** – NLMS

**Initialization**

1. Initialize parameter vector $\theta = \theta_0$. Set step size $\mu$ and regularization parameter $\varepsilon$

**For each time step k:**

2. Update parameter in NLMS form -

$$\theta(k+1) = \theta(k) + \frac{\mu}{x(k)^T x(k) + \varepsilon} e(k)x(k)$$

where $e(k) = f(k+1) - \phi(k)\theta(k)$ is the residual and $x(k)^T x(k) + \varepsilon$ is the normalization to avoid negligible $x(k) = [f'(k)\ dP(k)]$ such that the error is over-amplified.

**Algorithm 6** – FF-RLS

**Initialization**

1. Set forgetting factor $\lambda \in (0,1)$
2. Initialize parameter vector $\theta_0 = [\alpha_0; \gamma_0]$ with $\gamma_0 = 1/2H_0$. Initialize covariance matrix $P_0$ according to the scale of the parameter and its possible change.

**For each time step k:**

1. Compute time prediction –

$$\hat{y}_k = \phi_k \theta_{k-1}, \qquad \text{with } \phi_k = [f'_k\ \ dP_k\ dt]$$

2. Compute error –

$$e_k = f'_{k+1} - \hat{y}_k$$

3. Compute Kalman Gain –

$$K_k = \frac{P_{k-1}\phi_k'}{\lambda + \phi_k P_{k-1}\phi_k'}$$

4. Update parameter estimate –

$$\theta_k = \theta_{k-1} + K_k e_k, \qquad \theta_k \in [\theta_{\min}, \theta_{\max}]$$

5. Update Covariance –

$$P_k = \frac{1}{\lambda}(I - K_k\phi_k)P_{k-1}$$

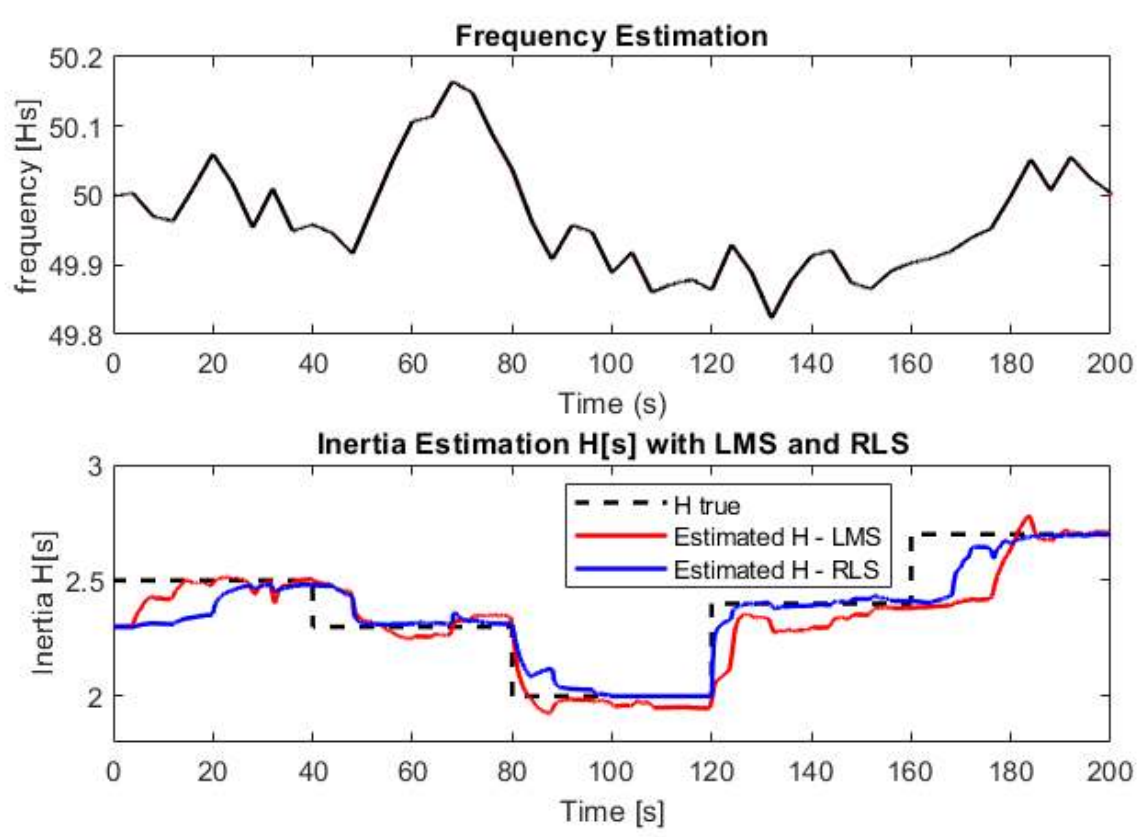

**Fig. 5**. Estimated Inertia with NLMS and RLS

**Table 5** - Performance Index for Inertia Estimation - Parameter Estimation with NLMS and RLS Algorithm

| H(0) | Method | MSE (Mean ± S.D.) | Max Error (Mean ± S.D.) | Avg H (Mean) | Bias (Mean) |
|---|---|---|---|---|---|
| 2.7 | LMS | 0.042 ± 0.030 | 0.47 ± 0.09 | 2.34 | -0.04 |
| | RLS | 0.009 ± 0.003 | 0.39 ± 0.01 | 2.38 | -0.00 |
| 2.8 | LMS | 0.070 ± 0.035 | 0.58 ± 0.11 | 2.27 | -0.11 |
| | RLS | 0.009 ± 0.003 | 0.39 ± 0.01 | 2.38 | +0.00 |
| 2.3 | LMS | 0.059 ± 0.030 | 0.51 ± 0.10 | 2.28 | -0.10 |
| | RLS | 0.012 ± 0.004 | 0.39 ± 0.01 | 2.35 | -0.03 |

## 5. Effect of Inertia Estimation in Inertia Extraction with Wind Generation - MPC

Inertia estimation, as discussed in **Section 1**, is often used to deduce ROCOF and frequency nadir under single largest contingency disturbance and to help avoid UFLS and passive loss-of-main protection activation in the current stage. Even more critical, regional inertia tracking is often required when inertia resources are spatially uneven which could possibly lead to localized frequency oscillations or swings such that inertia monitoring helps identify weak areas [10]-[12]. In the future, inertia forecast helps reserves scheduling and dispatch with IBRs and BESSs under market operations as an ancillary service. Also, fast frequency response (FFRs) allocation and placement as a frequency activation or adaptive control to release inertia from the IBRs are required to secure or arrest the grid from frequency disturbance. After a proper inertia estimation in the grid at the point of common coupling (PCC) of a wind farm, the wind farm can activate a model predictive control (MPC) to release inertia optimally as an additional feature against the limited or expensive output from other IBRs or HVDC.

As a dilemma, if the MPC is effective to eliminate or limit all possible frequency events, the IEAs cannot be effective to estimate the inertia without persistent excitation. Hence, all estimations must be performed under maximum power point (MPP) or virtual synchronous generator (VSG) mode of operation such that the power setpoint is kept constant and all other power deviation is from the external grid.

Lai & Hou (2025) suggested the first MPC controller with the inertia extraction from wind generations with dynamics in (6a) such that both 2-norm of system frequency deviation and the rotational speed of wind turbine (as a parameter on the torque balance between available wind torque and electrical torque driven by the generator side converter) were minimized as (6b) to obtain an optimal solution [42]. As discussed, it is impossible to have a faster controller than the observer (or estimator) to estimate the inertia constant and clear off all frequency deviation simultaneously.

Hence, a proper solution to improve the algorithm, as depicted in **Algorithm 7**, suggested by Lai & Hou (2025) is a gain-scheduled quadratic programming MPC (GS-QP-MPC) version, which is to estimate the inertia constant when the MPC model is often de-activated and choose the MPC evolution model with the inertia estimate to obtain an optimal input.

**Linearized Dynamics -**

$$\begin{cases} \dfrac{d\Omega}{dt} = \dfrac{1}{J}\left(\dfrac{\partial T_M(\Omega)}{\partial \Omega}|_0 \Omega - \Delta T_E\right) \\ \dfrac{d\omega}{dt} = \dfrac{\omega_B T_{E0}}{2HP_B}\Omega + \dfrac{\omega_B \Omega_0}{2HP_B}\Delta T_E - \dfrac{\omega_B}{2H}\dfrac{\Delta P_L}{P_B} \end{cases} \tag{6a}$$

where $\Omega = \Omega^* - \Omega_0$ and $\omega$ are the linearized shifted rotational speed of turbine and system frequency respectively; $\Delta T_E = T_E - T_{E0}$ is the electrical torque due to inverter control for real power output; $\Delta P_L$ is the additional load disturbance. $T_{E0}$, $\Omega_0$ and $P_B$ are the initial torque [N], initial rotational speed [rad/s] and power base [MVA] respectively.

**Stack Optimal Cost for MPC -**

$$x_{k+1} = \begin{cases} Ax_k + Bu_k & k \in [0, N-1] \\ (A + BK_{LQ})x_k & k \in [N, \infty) \end{cases} \tag{6b}$$

$$J(\mathbf{u}) = \frac{1}{2}\mathbf{x}_k^T\mathbf{Q}\mathbf{x}_k + \mathbf{u}_k^T\mathbf{R}\mathbf{u} + \mathbf{x}_N^T\mathbf{S}\mathbf{x}_N, \qquad (\mathbf{x}_k, \mathbf{u}_k) \in \mathbb{Z}$$

where $x_k = [\Omega \;\; \omega]^T, u_k, K_{LQ}, \mathbf{x}_k = [x_k \;\; x_{k-1} \dots x_{k-N+1}], \mathbf{Q}, \mathbf{R}, \mathbf{S}$ are the discretized state, input at time k, LQ optimal gain for the problem, stacked state at time k, cost associated with state, input and terminal state respectively.

**Algorithm 7** – GS-QP-MPC

**MPC Inactive (f'< $f_0$)**

1. Estimate $H$ with RLS algorithm (**Algorithm 6**) and store it as $H^*$

**MPC Active (f' > $f_0$ or ROCOF > $ROCOF_0$ or dP > $dP_0$)** [42]

2. Apply Forward Euler to linearize system matrix with $H^*$ and as the initial point $x_0$.

3. Determine the stacked dynamics in quadratic form as suggested in [42] and apply quadratic programming solver to determine the input $\mathbf{u}$
4. Apply the first input $\mathbf{u}[1]$ for $t$ seconds.
5. Return to Step 1 or Step 2 according to the requirement

The simulation is an infinite sink which does not change the inertia value upon the control activation of wind generators at the PCC connected to an aggregated wind farm with output 746MW. Small power steps deviation of 0.005pu has been introduced as normal load fluctuations to create frequency change as persistent excitation over time. At t = 50s, a generator with large inertia yet low load trips (-0.08pu) such that it creates a jump in inertia constant. The LMS and RLS algorithm is to learn the inertia constant when MPC is de-activated. At t = 220s, the frequency event occurs and an active loss-of-main protection signal, as an event start signal, activates the GS-QP-MPC to arrest the frequency.

From the simulation result as depicted in **Figure 6**, the inertia estimates with LMS and RLS algorithm are stable at t = 60s. The inertia estimate value, possibly rounded off to 2.5 to select the pre-defined state evolution model (over H = [2.3, 2.5, 2.7, 2.9, 3.0]) such that the state-evolution model with H = 2.5 will be used in **Algorithm 7**, instead of a fixed value of H = 2.8 over time. In **Figure 7**, various parameters in the wind farm interconnection including the rotation speed of the turbines, system frequency, torque setpoint and power setpoint to be activated by grid forming converter (GFM) is shown and the algorithm (GS-QP-MPC) is compared with the base case without any frequency response, PD controller, VSG controller which estimates the power deviation according to the ROCOF, and the QP-MPC with incorrect H.

Under the frequency event, PD controller responds and releases the frequency reserves according to the frequency deviation and ROCOF with the Kp and Kd parameters; while other controllers including VSG, QP-MPC and GS-QP-MPC drive optimally to release the inertia to support the grid. It indicates the difference between the expectation in general grid codes which require the wind farm to be possibly capable of releasing inertia with a droop value or an inertia constant and the actual nonlinearity possibly driven by the power controllers.

It is noted that the power output in **Figure 7(d)** for QP-MPC and GS-QP-MPC are close as both controllers can secure the frequency change. In **Figure 8**, however, the torque setpoint for QP-MPC (incorrect H = 2.8) and GS-QP-MPC (closer H = 2.5 updated by the RLS parameter estimator) under the true inertia constant value at H = [2.5, 2.3]s at t = [200, 250]s is close but QP-MPC setpoint has a small oscillation, as the incorrect H underestimates the ROCOF and release less output; and after a certain time period it discovers that the actual ROCOF goes much larger and it requires to release much more to secure until it discovers itself in the opposite condition. Similarly, with the same cost function as illustrated in **Figure 9**, the returned cost for the GS-QP-MPC is consistently smaller than that of the QP-MPC (H = 2.8), indicating that GS-QP-MPC is a better suboptimal solution due to the parameter mismatch. Both monotonically decrease over time as the QP-MPC guarantees.

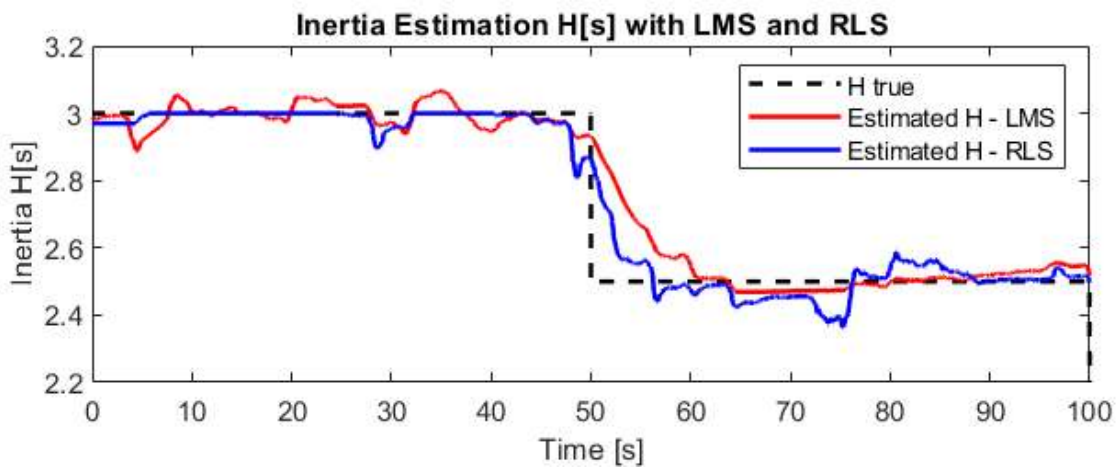

**Fig. 6**. Estimated Inertia (NLMS and RLS) in 0 – 100s

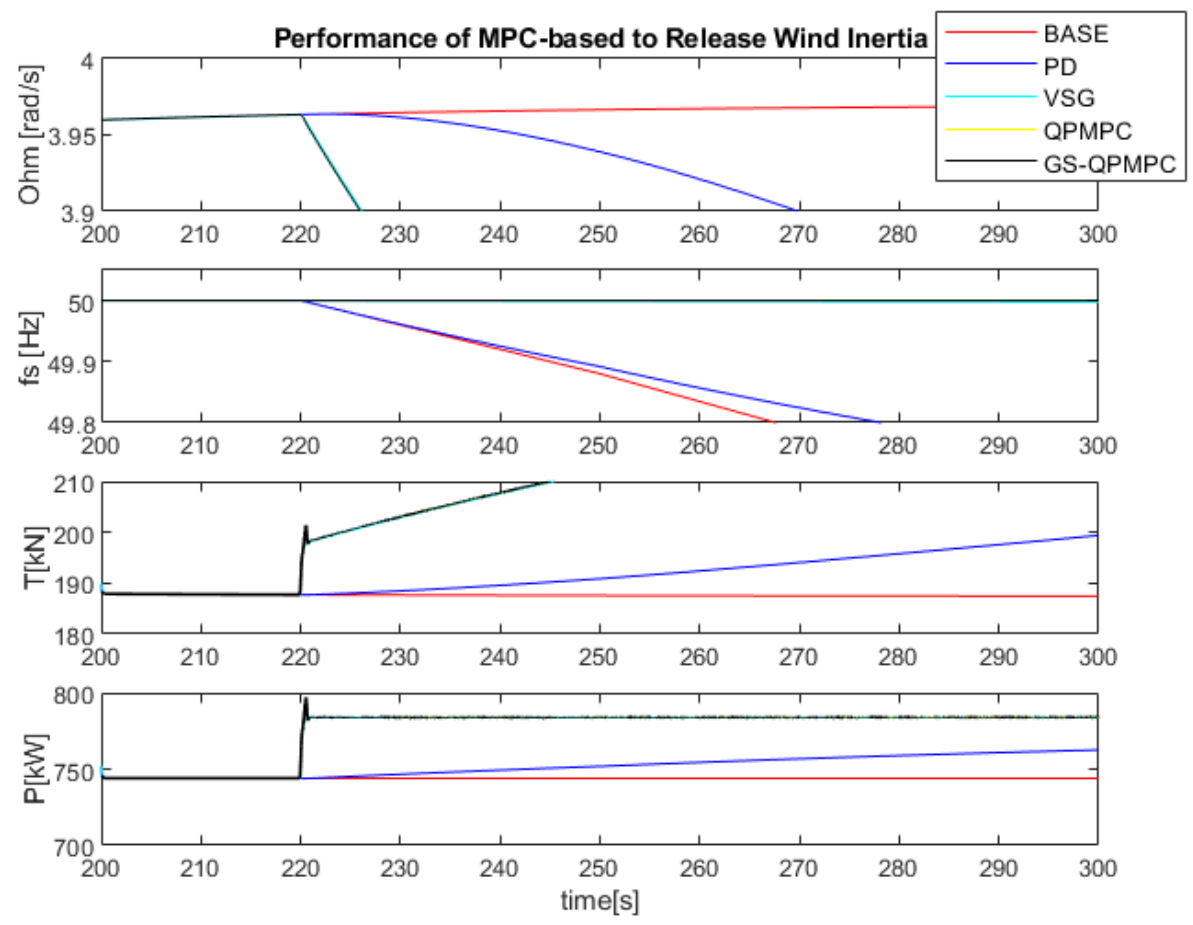

**Fig. 7**. (a) Turbine Rotation Speed; (b) System Frequency; (c) Electrical Torque and (d) Power Output upon a Power Step of 0.1pu

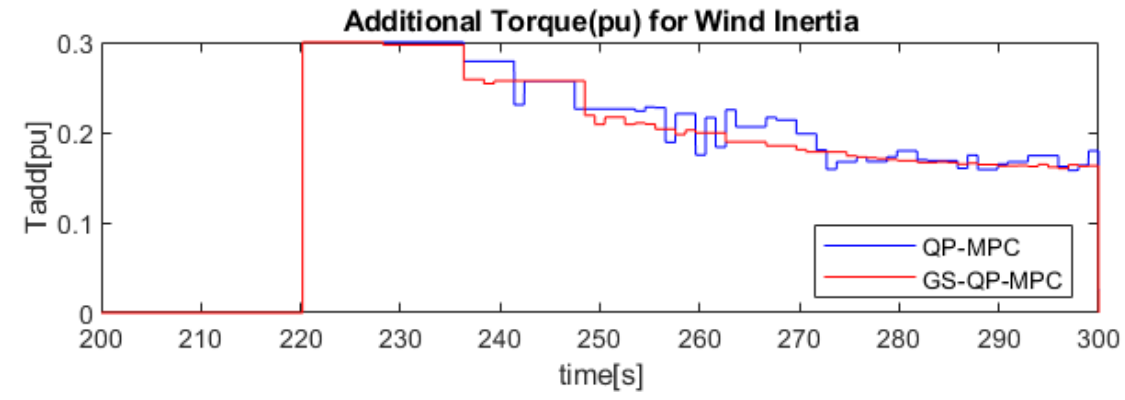

**Fig. 8**. Electrical Torque Setpoint with QP-MPC (H = 2.8) vs GS-QP-MPC (H = 2.5) upon True H = [2.5, 2.3]

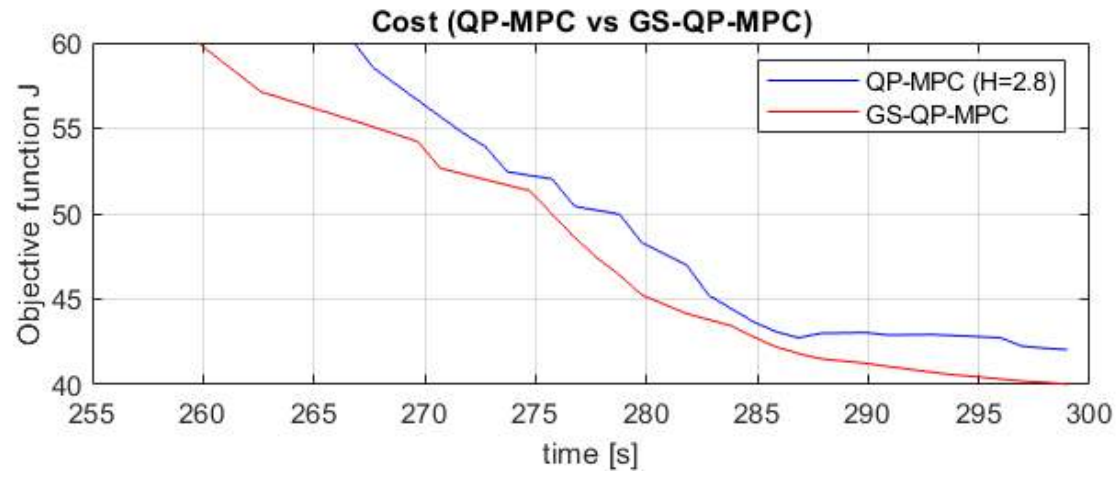

**Fig. 9**. Cost for QP-MPC (H = 2.8) and GS-QP-MPC (H = 2.5) upon True H = [2.5, 2.3]

## 6. Conclusion

This paper presents a comparative analysis on the inertia estimation algorithms typically used in the industries and academics, including measurement-based method, model-based method with KF which is based on stochastic model and minimum covariance, and data based method including LMS and RLS, with simulation results for comparison, and to introduce the current stage of inertia estimation applications in different utilities.

Results find that SR-KF and RLS algorithms provide the most accurate and speedy results on inertia estimation even without clipping, smoothening or other techniques to ensure its numerical stability or to avoid oscillation. The limitation in the simulation is that RLS is often prone to peaky fluctuations upon parameter jump and SR-KF requires parameter tuning to avoid sluggish performance. These are good solutions for frequency control and activation purposes. However, for monitoring and dispatch purposes in which speed is not a concern and smoothened result is much preferred, UKF and MMAE or measurement-based solution with polynomial fitting is a good alternative. LMS with a smaller step $\mu$ is another choice.

This paper also introduces a brand-new Gain Scheduling – Quadratic Programming based Model Predictive Control (GS-QP-MPC) to allow change in model parameters for a better objective function elimination and avoidance of oscillation in the torque input. Although a fixed but mismatch parameter model for QP-MPC can still provide a sub-optimal result, small but continuous oscillation should be eliminated to avoid high cycle fatigue for the wind turbine. The only limitation for such control is that it must have a deadband for the MPC such that the estimator can perform. The advantage of Kalman Filter and RLS solution over measurement-based approach is to avoid the calculation or determination of ROCOF and start-time identification.

In terms of parameter estimation, more studies should be performed on tune-free estimators with Kalman Filter, LMS and RLS. Also, the current studies have no outlier or faulty data such that the performance of the Kalman Filter and RLS estimators are still reliable. The combination of estimators at different locations should also be studied to provide a reliable and indicative inertia constant. For the application purpose, more studies should be made on the secondary power-frequency control loop to provide a proper setpoint to counteract any system dynamics. The assumption on linear relationships should be avoided as the frequency control for whole system is never a linear relationship under a large-signal instability.

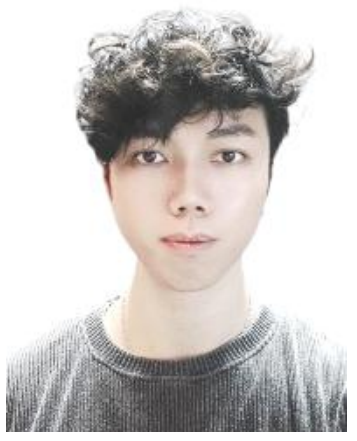

**Mr. Karl M. H. LAI**

Karl M.H. Lai received his MSc in HKU in 2022. In 2019, he joined CLP as a protection engineer specializing in system, generator and IBR protection and control. He has a particular interest in advanced control, filtering and estimation in IBR and HVDC systems.

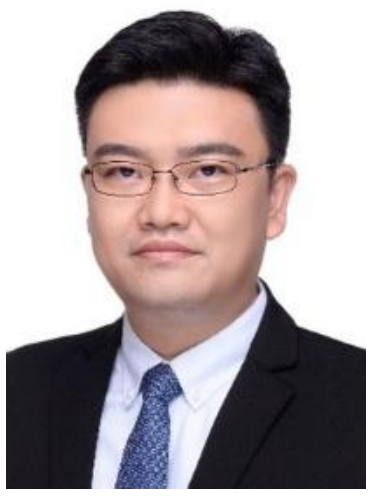

**Prof. Yunhe HOU**

Yunhe Hou is an Associate Professor with Department of Electrical and Computer Engineering, HKU. He received his PhD in 2005 from Huazhong University of Science and Technology, China. He specializes in power system resilience and planning and optimization in power systems.

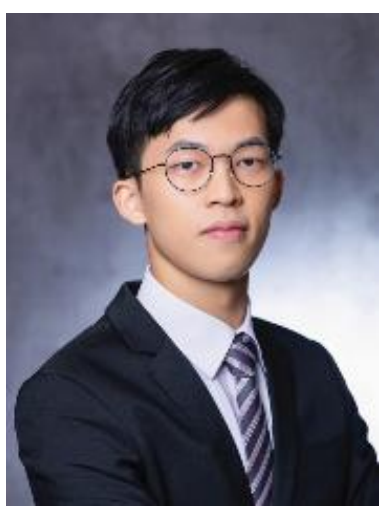

**Mr. Kwunhang WONG**

Kwunhang Wong is a PhD candidate in Department of Electrical and Computer Engineering, HKU. He specializes in neuromorphic computing, hardware-software co-design and differential privacy.